\documentclass[11pt]{article}

\usepackage{hyperref}
\usepackage{amssymb}
\usepackage{bbold}

\title{\bf GMCALC: a calculator for the Georgi-Machacek model\footnote{Code available from http://people.physics.carleton.ca/$\sim$logan/gmcalc/ .}}

\author{Katy Hartling\thanks{\tt khally@physics.carleton.ca}, \ Kunal Kumar\thanks{\tt kkumar@physics.carleton.ca}, \ and Heather E.\ Logan\thanks{\tt logan@physics.carleton.ca} \\
{\it Ottawa-Carleton Institute for Physics, Carleton University,} \\ 
{\it 1125 Colonel By Drive, Ottawa K1S 5B6 Canada}}

\date{Version 1.0: December 20, 2014}

\begin{document}
\maketitle

\begin{abstract}
The Georgi-Machacek model adds scalar triplets to the Standard Model Higgs sector in such a way as to preserve custodial SU(2) symmetry in the scalar potential. This allows the triplets to have a non-negligible vacuum expectation value while satisfying constraints from the $\rho$ parameter. Depending on the parameters, the 125~GeV neutral Higgs particle can have couplings to $WW$ and $ZZ$ larger than in the Standard Model due to mixing with the triplets. The model also contains singly- and doubly-charged Higgs particles that couple to vector boson pairs at tree level ($WZ$ and like-sign $WW$, respectively).

GMCALC is a self-contained FORTRAN program that, given a set of input parameters, calculates the particle spectrum and tree-level couplings in the Georgi-Machacek model, checks theoretical and indirect constraints, and computes the branching ratios and total widths of the scalars. It also generates a param\_card.dat file for MadGraph5 to be used with the corresponding FeynRules model implementation. 
\end{abstract}

\newpage
\tableofcontents
\newpage

\section{Introduction}

The Georgi-Machacek (GM) model~\cite{Georgi:1985nv,Chanowitz:1985ug} is an extension of the Standard Model (SM) Higgs sector containing additional scalars in the triplet representation of SU(2)$_L$.  The particle content is such that an additional global SU(2)$_R$ symmetry can be imposed by hand on the scalar potential.  This ensures that the custodial SU(2) symmetry, which fixes $\rho \equiv M_W^2/M_Z^2 \cos^2\theta_W = 1$ at tree level in the SM, is preserved after electroweak symmetry breaking.  

Without the stringent constraint from the $\rho$ parameter, the vacuum expectation value (vev) of the triplets can be large, leading to interesting phenomenology.  In particular, depending on the parameters, the 125~GeV neutral Higgs particle can have couplings to $WW$ and $ZZ$ larger than in the SM due to mixing with the triplets. The model also contains singly- and doubly-charged Higgs particles that couple to vector boson pairs at tree level, leading to $H_5^+ \to W^+Z$ and like-sign $H_5^{++} \to W^+ W^+$ signatures.  Such an $H^+ W^- Z$ coupling is absent at tree level in two Higgs doublet models (2HDMs), and the $H^{++} W^- W^-$ coupling is severely suppressed in triplet models without custodial symmetry in which the triplet vev is forced to be very small by the experimental constraint from the $\rho$ parameter.

This manual describes the FORTRAN code GMCALC.  Given a set of model parameters, GMCALC calculates the mass spectrum and relevant mixing angles in the scalar sector, as well as the tree-level couplings of the scalars.  It also checks that theoretical constraints from perturbative unitarity of the quartic scalar couplings, bounded-from-belowness of the scalar potential, and the absence of deeper custodial-symmetry-breaking minima are satisfied.  The code also checks consistency of the parameter point with indirect experimental constraints from the $S$ parameter, $b \to s \gamma$, and $B_s^0 \to \mu^+\mu^-$.  Finally, it computes the branching ratios and total widths of the scalars.  Most of the code is based on our work in Refs.~\cite{HKL,indirect}.

GMCALC includes a routine to generate a param\_card.dat file for MadGraph5 to be used with the corresponding FeynRules model implementation.  The FeynRules implementation for the Georgi-Machacek model can be downloaded from the model database at http://feynrules.irmp.ucl.ac.be.

This manual is organized as follows.  In Sec.~\ref{sec:model} we give a brief description of the GM model and set our notation.  In Sec.~\ref{sec:thy} we review the theoretical constraints and their implementation.  In Sec.~\ref{sec:indir} we describe the indirect experimental constraints that are implemented in the code.  In Sec.~\ref{sec:decays} we summarize the computation of the decay partial widths of the scalars and specify the approximations made in the code.  Finally in Sec.~\ref{sec:using} we give instructions for using the GMCALC code.

\section{Georgi-Machacek model}
\label{sec:model}

\subsection{Scalar potential}

The scalar sector of the Georgi-Machacek model consists of the usual complex doublet $(\phi^+,\phi^0)$ with hypercharge\footnote{We use $Q = T^3 + Y/2$.} $Y = 1$, a real 
triplet $(\xi^+,\xi^0,\xi^-)$ with $Y = 0$, and  a complex triplet $(\chi^{++},\chi^+,\chi^0)$ with $Y=2$.  The doublet is responsible for the fermion masses as in the SM.
In order to make the global SU(2)$_L \times$SU(2)$_R$ symmetry explicit, we write the doublet in the form of a bi-doublet $\Phi$ and combine the triplets to form a bi-triplet $X$:
\begin{eqnarray}
	\Phi &=& \left( \begin{array}{cc}
	\phi^{0*} &\phi^+  \\
	-\phi^{+*} & \phi^0  \end{array} \right), \\
	X &=&
	\left(
	\begin{array}{ccc}
	\chi^{0*} & \xi^+ & \chi^{++} \\
	 -\chi^{+*} & \xi^{0} & \chi^+ \\
	 \chi^{++*} & -\xi^{+*} & \chi^0  
	\end{array}
	\right).
	\label{eq:PX}
\end{eqnarray}
The vevs are defined by $\langle \Phi  \rangle = \frac{ v_{\phi}}{\sqrt{2}} \mathbb{1}_{2\times2}$  and $\langle X \rangle = v_{\chi} \mathbb{1}_{3 \times 3}$, where the Fermi constant constrains
\begin{equation}
	v_{\phi}^2 + 8 v_{\chi}^2 \equiv v^2 = \frac{1}{\sqrt{2} G_F} \approx (246~{\rm GeV})^2.
	\label{eq:vevrelation}
\end{equation} 
Note that the two triplet fields $\chi^0$ and $\xi^0$ must obtain the same vev in order to preserve custodial SU(2).
Furthermore we will decompose the neutral fields into real and imaginary parts according to
\begin{equation}
	\phi^0 \to \frac{v_{\phi}}{\sqrt{2}} + \frac{\phi^{0,r} + i \phi^{0,i}}{\sqrt{2}},
	\qquad
	\chi^0 \to v_{\chi} + \frac{\chi^{0,r} + i \chi^{0,i}}{\sqrt{2}}, 
	\qquad
	\xi^0 \to v_{\chi} + \xi^0,
\end{equation}
where we note that $\xi^0$ is already a real field.

Using the notation of Ref.~\cite{HKL}, the most general gauge-invariant scalar potential involving these fields that conserves custodial SU(2) is given by
\begin{eqnarray}
	V(\Phi,X) &= & \frac{\mu_2^2}{2} {\rm Tr}(\Phi^\dagger \Phi) 
	+  \frac{\mu_3^2}{2}  {\rm Tr}(X^\dagger X)  
	+ \lambda_1 [{\rm Tr}(\Phi^\dagger \Phi)]^2  
	+ \lambda_2 {\rm Tr}(\Phi^\dagger \Phi) {\rm Tr}(X^\dagger X)   \nonumber \\
          & & + \lambda_3 {\rm Tr}(X^\dagger X X^\dagger X)  
          + \lambda_4 [{\rm Tr}(X^\dagger X)]^2 
           - \lambda_5 {\rm Tr}( \Phi^\dagger \tau^a \Phi \tau^b) {\rm Tr}( X^\dagger t^a X t^b) 
           \nonumber \\
           & & - M_1 {\rm Tr}(\Phi^\dagger \tau^a \Phi \tau^b)(U X U^\dagger)_{ab}  
           -  M_2 {\rm Tr}(X^\dagger t^a X t^b)(U X U^\dagger)_{ab}.
           \label{eq:potential}
\end{eqnarray} 
(A translation table to other notations used in the literature is given in the appendix of Ref.~\cite{HKL}.)
Here the SU(2) generators for the doublet representation are $\tau^a = \sigma^a/2$ with $\sigma^a$ being the Pauli matrices,
the generators for the triplet representation are
\begin{equation}
	t^1= \frac{1}{\sqrt{2}} \left( \begin{array}{ccc}
	 0 & 1  & 0  \\
	  1 & 0  & 1  \\
	  0 & 1  & 0 \end{array} \right), \quad  
	  t^2= \frac{1}{\sqrt{2}} \left( \begin{array}{ccc}
	 0 & -i  & 0  \\
	  i & 0  & -i  \\
	  0 & i  & 0 \end{array} \right), \quad 
	t^3= \left( \begin{array}{ccc}
	 1 & 0  & 0  \\
	  0 & 0  & 0  \\
	  0 & 0 & -1 \end{array} \right),
\end{equation}
and the matrix $U$, which rotates $X$ into the Cartesian basis, is given by
\begin{equation}
	 U = \left( \begin{array}{ccc}
	- \frac{1}{\sqrt{2}} & 0 &  \frac{1}{\sqrt{2}} \\
	 - \frac{i}{\sqrt{2}} & 0  &   - \frac{i}{\sqrt{2}} \\
	   0 & 1 & 0 \end{array} \right).
	 \label{eq:U}
\end{equation}
We note that all the operators in Eq.~(\ref{eq:potential}) are manifestly Hermitian, so that the parameters in the scalar potential must all be real.  Explicit CP violation is thus not possible in the Georgi-Machacek model.  

\subsection{Electroweak symmetry breaking and physical spectrum}

Minimizing the scalar potential yields the following constraints:
\begin{eqnarray}
	0 = \frac{\partial V}{\partial v_{\phi}} &=& 
	v_{\phi} \left[ \mu_2^2 + 4 \lambda_1 v_{\phi}^2 
	+ 3 \left( 2 \lambda_2 - \lambda_5 \right) v_{\chi}^2 - \frac{3}{2} M_1 v_{\chi} \right], 
		\label{eq:phimincond} \\
	0 = \frac{\partial V}{\partial v_{\chi}} &=& 
	3 \mu_3^2 v_{\chi} + 3 \left( 2 \lambda_2 - \lambda_5 \right) v_{\phi}^2 v_{\chi}
	+ 12 \left( \lambda_3 + 3 \lambda_4 \right) v_{\chi}^3
	- \frac{3}{4} M_1 v_{\phi}^2 - 18 M_2 v_{\chi}^2.
	\label{eq:chimincond}
\end{eqnarray}
Inserting $v_{\phi}^2 = v^2 - 8 v_{\chi}^2$ [Eq.~(\ref{eq:vevrelation})] into Eq.~(\ref{eq:chimincond}) yields a cubic equation for $v_{\chi}$ in terms of $v$, $\mu_3^2$, $\lambda_2$, $\lambda_3$, $\lambda_4$, $\lambda_5$, $M_1$, and $M_2$.  With $v_{\chi}$ (and hence $v_{\phi}$) in hand, Eq.~(\ref{eq:phimincond}) can be used to eliminate $\mu_2^2$ in terms of the parameters in the previous sentence together with $\lambda_1$.  We illustrate below how $\lambda_1$ can also be eliminated in favor of one of the custodial singlet Higgs masses $m_h$ or $m_H$ [see Eq.~(\ref{eq:lambda1})].

The physical field content is as follows.  The Goldstone bosons are given by
\begin{eqnarray}
	G^+ &=& c_H \phi^+ + s_H \frac{\left(\chi^++\xi^+\right)}{\sqrt{2}}, \nonumber\\
	G^0  &=& c_H \phi^{0,i} + s_H \chi^{0,i},
\end{eqnarray}
where
\begin{equation}
	c_H \equiv \cos\theta_H = \frac{v_{\phi}}{v}, \qquad
	s_H \equiv \sin\theta_H = \frac{2\sqrt{2}\,v_\chi}{v}.
\end{equation}
The physical fields can be organized by their transformation properties under the custodial SU(2) symmetry into a fiveplet, a triplet, and two singlets.  The fiveplet and triplet states are given by
\begin{eqnarray}
	H_5^{++}  &=&  \chi^{++}, \nonumber\\
	H_5^+ &=& \frac{\left(\chi^+ - \xi^+\right)}{\sqrt{2}}, \nonumber\\
	H_5^0 &=& \sqrt{\frac{2}{3}} \xi^0 - \sqrt{\frac{1}{3}} \chi^{0,r}, \nonumber\\
	H_3^+ &=& - s_H \phi^+ + c_H \frac{\left(\chi^++\xi^+\right)}{\sqrt{2}}, \nonumber\\
	H_3^0 &=& - s_H \phi^{0,i} + c_H \chi^{0,i}.
\end{eqnarray}
Within each custodial multiplet, the masses are degenerate at tree level.  Using Eqs.~(\ref{eq:phimincond}--\ref{eq:chimincond}) to eliminate $\mu_2^2$ and $\mu_3^2$, the fiveplet and triplet masses can be written as
\begin{eqnarray}
	m_5^2 &=& \frac{M_1}{4 v_{\chi}} v_\phi^2 + 12 M_2 v_{\chi} 
	+ \frac{3}{2} \lambda_5 v_{\phi}^2 + 8 \lambda_3 v_{\chi}^2, \nonumber \\
	m_3^2 &=&  \frac{M_1}{4 v_{\chi}} (v_\phi^2 + 8 v_{\chi}^2) 
	+ \frac{\lambda_5}{2} (v_{\phi}^2 + 8 v_{\chi}^2) 
	= \left(  \frac{M_1}{4 v_{\chi}} + \frac{\lambda_5}{2} \right) v^2.
\end{eqnarray}
Note that the ratio $M_1/v_{\chi}$ is finite in the limit $v_{\chi} \to 0$, as can be seen from Eq.~(\ref{eq:chimincond}) which yields
\begin{equation}
	\frac{M_1}{v_{\chi}} = \frac{4}{v_{\phi}^2} 
	\left[ \mu_3^2 + (2 \lambda_2 - \lambda_5) v_{\phi}^2 
	+ 4(\lambda_3 + 3 \lambda_4) v_{\chi}^2 - 6 M_2 v_{\chi} \right].
\end{equation}

The two custodial SU(2) singlets are given in the gauge basis by
\begin{eqnarray}
	H_1^0 &=& \phi^{0,r}, \nonumber \\
	H_1^{0 \prime} &=& \sqrt{\frac{1}{3}} \xi^0 + \sqrt{\frac{2}{3}} \chi^{0,r}.
\end{eqnarray}
These states mix by an angle $\alpha$ to form the two custodial-singlet mass eigenstates $h$ and $H$, defined such that $m_h < m_H$:
\begin{eqnarray}
	h &=& \cos \alpha \, H_1^0 - \sin \alpha \, H_1^{0\prime},   \\ \nonumber 
	H &=& \sin \alpha \, H_1^0 + \cos \alpha \, H_1^{0\prime}.
\end{eqnarray}
The mixing is controlled by the $2\times 2$ mass-squared matrix
\begin{equation}
	\mathcal{M}^2 = \left( \begin{array}{cc}
			\mathcal{M}_{11}^2 & \mathcal{M}_{12}^2 \\
			\mathcal{M}_{12}^2 & \mathcal{M}_{22}^2 \end{array} \right),
\end{equation}
where
\begin{eqnarray}
	\mathcal{M}_{11}^2 &=& 8 \lambda_1 v_{\phi}^2, \nonumber \\
	\mathcal{M}_{12}^2 &=& \frac{\sqrt{3}}{2} v_{\phi} 
	\left[ - M_1 + 4 \left(2 \lambda_2 - \lambda_5 \right) v_{\chi} \right], \nonumber \\
	\mathcal{M}_{22}^2 &=& \frac{M_1 v_{\phi}^2}{4 v_{\chi}} - 6 M_2 v_{\chi} 
	+ 8 \left( \lambda_3 + 3 \lambda_4 \right) v_{\chi}^2.
\end{eqnarray}
The mixing angle is fixed by 
\begin{eqnarray}
	\sin 2 \alpha &=&  \frac{2 \mathcal{M}^2_{12}}{m_H^2 - m_h^2},    \nonumber  \\
	\cos 2 \alpha &=&  \frac{ \mathcal{M}^2_{22} - \mathcal{M}^2_{11}  }{m_H^2 - m_h^2},    
\end{eqnarray}
and is chosen to be in the range $\alpha \in (-\pi/2, \pi/2]$, so that $\cos\alpha \geq 0$.  The masses are given by
\begin{eqnarray}
	m^2_{h,H} &=& \frac{1}{2} \left[ \mathcal{M}_{11}^2 + \mathcal{M}_{22}^2
	\mp \sqrt{\left( \mathcal{M}_{11}^2 - \mathcal{M}_{22}^2 \right)^2 
	+ 4 \left( \mathcal{M}_{12}^2 \right)^2} \right].
	\label{eq:hmass}
\end{eqnarray}

It is convenient to use the measured mass of the observed SM-like Higgs boson as an input parameter.  The coupling $\lambda_1$ can be eliminated in favor of this mass by inverting Eq.~(\ref{eq:hmass}):
\begin{equation}
	\lambda_1 = \frac{1}{8 v_{\phi}^2} \left[ m_h^2 
	+ \frac{\left( \mathcal{M}_{12}^2 \right)^2}{\mathcal{M}_{22}^2 - m_h^2} \right].
	\label{eq:lambda1}
\end{equation}
Note that in deriving this expression for $\lambda_1$, the distinction between $m_h$ and $m_H$ is lost.  This means that, depending on the values of $\mu_3^2$ and the other parameters, this (unique) solution for $\lambda_1$ will correspond to either the lighter or the heavier custodial singlet having a mass equal to the observed SM-like Higgs mass.

\subsection{Yukawa sector}

Fermion masses are generated through couplings to the complex doublet $\phi \equiv (\phi^+,\phi^0)$ in the same was as in the SM.  We neglect neutrino masses.  The relevant Lagrangian terms are
\begin{equation}
	\mathcal{L} \supset - \sum_{i=1}^3 \sum_{j=1}^3
	\left[ y^u_{ij} \bar u_{Ri} \tilde \phi^{\dagger} Q_{Lj}
	+ y^d_{ij} \bar d_{Ri} \phi^{\dagger} Q_{Lj} \right] 
	+ y^{\ell}_i \bar \ell_{Ri} \phi^{\dagger} L_{Li}
	+ {\rm h.c.},
\end{equation}
where $i,j$ run over the three generations and $\tilde \phi \equiv i \sigma^2 \phi^*$.  
The custodial singlets and triplet contain an admixture of $\phi$, and so couple to fermions.  The custodial fiveplet states do not couple to fermions.

The Feynman rules for neutral scalars coupling to fermion pairs are given as follows:
\begin{eqnarray}
	h \bar f f: &\quad& -i \frac{m_f}{v} \frac{\cos \alpha}{\cos \theta_H}, \qquad \qquad
	H \bar f f: \quad -i \frac{m_f}{v} \frac{\sin \alpha}{\cos \theta_H}, \nonumber \\
	H_3^0 \bar u u: &\quad& \frac{m_u}{v} \tan \theta_H \gamma_5, \qquad \qquad
	H_3^0 \bar d d: \quad -\frac{m_d}{v} \tan \theta_H \gamma_5.
\end{eqnarray}
Here $f$ denotes any charged fermion, $u$ stands for any up-type quark, and $d$ stands for any down-type quark or charged lepton.  

The Feynman rules for the vertices involving a charged scalar and two fermions are given as follows, with all particles incoming:
\begin{eqnarray}
	H_3^+ \bar u d: &\quad& -i \sqrt{2} V_{ud} \tan\theta_H
		\left( \frac{m_u}{v} P_L - \frac{m_d}{v} P_R \right), \nonumber \\
	H_3^{+*} \bar d u: &\quad& -i \sqrt{2} V_{ud}^* \tan\theta_H 
		\left( \frac{m_u}{v} P_R - \frac{m_d}{v} P_L \right), \nonumber \\
	H_3^+ \bar \nu \ell: &\quad& i \sqrt{2} \tan\theta_H \frac{m_{\ell}}{v} P_R, \nonumber \\
	H_3^{+*} \bar \ell \nu: &\quad& i \sqrt{2} \tan\theta_H \frac{m_{\ell}}{v} P_L.
\end{eqnarray}
Here $V_{ud}$ is the appropriate element of the Cabibbo-Kobayashi-Maskawa matrix and the projection operators are defined as $P_{R,L} = (1 \pm \gamma_5)/2$.

\section{Theoretical constraints}
\label{sec:thy}

\subsection{Tree-level unitarity}

We implement the conditions for unitarity of tree-level $2\to 2$ scalar particle scattering amplitudes computed in Refs.~\cite{Aoki:2007ah,HKL}.  These were computed by imposing $|{\rm Re}\, a_0| < 1/2$ on the eigenvalues of the zeroth partial wave amplitude coupled-channel matrix, and read
\begin{eqnarray}
	\sqrt{ \left( 6 \lambda_1 - 7 \lambda_3 - 11 \lambda_4 \right)^2 + 36 \lambda_2^2}
	+ \left| 6 \lambda_1 + 7 \lambda_3 + 11 \lambda_4 \right| &<& 4 \pi, \nonumber \\
	\sqrt{ \left( 2 \lambda_1 + \lambda_3 - 2 \lambda_4 \right)^2 + \lambda_5^2}
	+ \left| 2 \lambda_1 - \lambda_3 + 2 \lambda_4 \right| &<& 4 \pi, \nonumber \\
	\left| 2 \lambda_3 + \lambda_4 \right| &<& \pi, \nonumber \\
	\left| \lambda_2 - \lambda_5 \right| &<& 2 \pi.
	\label{eq:uni}
\end{eqnarray}

\subsection{Bounded-from-below requirement on the potential}

We implement the conditions that ensure the scalar potential is bounded from below as computed in Ref.~\cite{HKL}.  They read as follows:
\begin{eqnarray}
	\lambda_1 &>& 0, \nonumber \\
	\lambda_4 &>& \left\{ \begin{array}{l l}
		- \frac{1}{3} \lambda_3 & {\rm for} \ \lambda_3 \geq 0, \\
		- \lambda_3 & {\rm for} \ \lambda_3 < 0, \end{array} \right. \nonumber \\
	\lambda_2 &>& \left\{ \begin{array}{l l}
		\frac{1}{2} \lambda_5 - 2 \sqrt{\lambda_1 \left( \frac{1}{3} \lambda_3 + \lambda_4 \right)} &
			{\rm for} \ \lambda_5 \geq 0 \ {\rm and} \ \lambda_3 \geq 0, \\
		\omega_+(\zeta) \lambda_5 - 2 \sqrt{\lambda_1 ( \zeta \lambda_3 + \lambda_4)} &
			{\rm for} \ \lambda_5 \geq 0 \ {\rm and} \ \lambda_3 < 0, \\
		\omega_-(\zeta) \lambda_5 - 2 \sqrt{\lambda_1 (\zeta \lambda_3 + \lambda_4)} &
			{\rm for} \ \lambda_5 < 0, 
			\end{array} \right.
	\label{eq:bfbcond2}
\end{eqnarray}
where 
\begin{equation}
	\omega_{\pm}(\zeta) = \frac{1}{6}(1 - B) \pm \frac{\sqrt{2}}{3} \left[ (1 - B) \left(\frac{1}{2} + B\right)\right]^{1/2},
\end{equation}
with
\begin{equation}
	B \equiv \sqrt{\frac{3}{2}\left(\zeta - \frac{1}{3}\right)} \in [0,1].
\end{equation}
The last two conditions for $\lambda_2$ in Eq.~(\ref{eq:bfbcond2}) must be satisfied for all values of $\zeta \in \left[ \frac{1}{3}, 1 \right]$.  We implement this through a 1000-point scan over $\zeta$ in the specified range.

\subsection{Absence of deeper custodial symmetry-breaking minima}

Finally, we implement a check that the scalar potential possesses no custodial symmetry-breaking minima that are deeper than the desired custodial symmetry-preserving minimum, following the procedure described in Ref.~\cite{HKL}.  We write the scalar potential as
\begin{equation}
	V = \frac{\mu_2^2}{2} a^2 + \frac{\mu_3^2}{2} b^2 + \lambda_1 a^4 + \lambda_2 a^2 b^2
	+ \zeta \lambda_3 b^4 + \lambda_4 b^4 - \omega \lambda_5 a^2 b^2 - \sigma M_1 a^2 b 
	- \rho M_2 b^3,
\end{equation}
where $a^2 = {\rm Tr}(\Phi^{\dagger}\Phi)$ and $b^2 = {\rm Tr}(X^{\dagger}X)$ and the dimensionless coefficients $\zeta$, $\omega$, $\sigma$, and $\rho$ vary with varying triplet field configurations.  The minimum of $V$ is always traced out by the path~\cite{HKL}
\begin{eqnarray}
	\zeta &=& \frac{1}{2} \sin^4 \theta + \cos^4 \theta, \nonumber \\
	\omega &=& \frac{1}{4} \sin^2 \theta + \frac{1}{\sqrt{2}} \sin \theta \cos \theta, \nonumber \\
	\sigma &=& \frac{1}{2\sqrt{2}} \sin \theta + \frac{1}{4} \cos \theta, \nonumber \\
	\rho &=& 3 \sin^2 \theta \cos \theta,
	\label{eq:thetaparam}
\end{eqnarray}
with $\theta \in [0, 2\pi)$.  Our desired electroweak-breaking and custodial SU(2)-preserving vacuum corresponds to $\theta = \cos^{-1} (1/\sqrt{3})$.  The vacuum $\theta = \pi + \cos^{-1} (1/\sqrt{3})$ is also acceptable; it corresponds to negative $b$.  The depths of these vacua are determined by applying the minimization conditions and solving the resulting cubic and quadratic equations to determine the values of $a$ and $b$ that minimize the potential, then evaluating $V$ at this minimum.

This procedure is then repeated for other values of $\theta$ [corresponding to vacua that spontaneously break custodial SU(2)] using a 1000-point scan over $\theta \in [0, 2\pi)$.  Parameter points fail this check if any vacuum solution exists in which $V$ is lower than the value in the desired vacuum.

\section{Indirect experimental constraints}
\label{sec:indir}

Indirect constraints from the $S$ parameter, $b \to s \gamma$, and $B_s^0 \to \mu^+ \mu^-$ are implemented in the code.  A detailed physics description is given in Ref.~\cite{indirect}.  Currently the constraint from $b \to s \gamma$ is stronger than that from $B_s^0 \to \mu^+\mu^-$, but that may change in the next several years as more data is collected at the CERN Large Hadron Collider.

\subsection{$S$ parameter}

When the new physics is not light compared to $M_Z$, the $S$ parameter can be written in terms of the derivatives $\Pi^{\prime}(0) \equiv d \Pi(p^2)/d p^2 |_{p^2 = 0}$ of the gauge boson self-energies as
\begin{equation}
	S = \frac{4 s_W^2 c_W^2}{\alpha_{EM}} \left[\Pi'_{ZZ}(0)
	-\frac{c_W^2-s_W^2}{s_W c_W}\Pi'_{Z\gamma}(0)-\Pi'_{\gamma\gamma}(0)\right].
\end{equation}
The new physics contribution in the GM model, relative to the SM for a reference Higgs mass $m_h^{\rm SM}$, is~\cite{indirect}
\begin{eqnarray}
	S &=& \frac{s_W^2 c_W^2}{e^2\pi}
	\left\{-\frac{e^2}{12 s_W^2 c_W^2}\left(\log{m_3^2}+5\log{m_5^2}\right)
	+2|g_{ZhH_3^0}|^2\,f_1(m_h,m_3)\right. \nonumber\\
	&& \left.+ 2|g_{ZHH_3^0}|^2\,f_1(m_H,m_3)
	+2\left(|g_{ZH_5^0H_3^0}|^2+2|g_{ZH_5^+H_3^{+*}}|^2\right)f_1(m_5,m_3)\right. 
	\nonumber\\
	&&\left.+|g_{ZZh}|^2\left[\frac{f_1(M_Z,m_h)}{2 M_Z^2}-f_3(M_Z,m_h)\right]
	+|g_{ZZH}|^2\left[\frac{f_1(M_Z,m_H)}{2 M_Z^2}-f_3(M_Z,m_H)\right]\right. \nonumber\\
	&&\left.+|g_{ZZH_5^0}|^2\left[\frac{f_1(M_Z,m_5)}{2 M_Z^2}-f_3(M_Z,m_5)\right]\right. \nonumber\\
	&&\left.+2|g_{ZW^+H_5^{+*}}|^2\left[\frac{f_1(M_W,m_5)}{2 M_W^2}-f_3(M_W,m_5)\right]
	\right. \nonumber \\
	&&\left. -|g_{ZZh}^{\rm SM}|^2\left[\frac{f_1(M_Z,m_h^{\rm SM})}{2 M_Z^2}-f_3(M_Z,m_h^{\rm SM})\right] \right\},
	\label{eq:SGM}
\end{eqnarray}
where 
\begin{eqnarray}
	f_1(m_1,m_2) &=&
	\frac{5(m_2^6-m_1^6) + 27 (m_1^4 m_2^2-m_1^2 m_2^4) 
	+ 12 (m_1^6-3 m_1^4 m_2^2) \log m_1 
	+ 12(3 m_1^2 m_2^4-m_2^6)\log m_2}{36(m_1^2-m_2^2)^3}, \nonumber \\
	f_3(m_1,m_2) &=& \frac{m_1^4 - m_2^4 
	+ 2 m_1^2 m_2^2\left(\log m_2^2 - \log m_1^2\right)}{2 (m_1^2 - m_2^2 )^3}.
\end{eqnarray}
For numerical stability we use an expansion in $\epsilon \equiv \frac{m_2^2}{m_1^2} - 1$ when $m_1^2 \simeq m_2^2$ to within a part in $10^{-4}$, 
\begin{equation}
	f_1(m_1,m_2) \simeq \frac{1}{6} \log m_1^2 + \frac{\epsilon}{12}, \qquad \qquad
	f_3(m_1,m_2) \simeq \frac{1}{6 m_1^2} - \frac{\epsilon}{12 m_1^2}.
\end{equation}

The couplings that appear in Eq.~(\ref{eq:SGM}) are given by~\cite{HKL}
\begin{eqnarray}
	g_{ZhH_3^0} &=& -i\sqrt{\frac{2}{3}}\frac{e}{s_Wc_W}\left(s_\alpha \frac{v_\phi}{v} + \sqrt{3} c_\alpha \frac{v_\chi}{v} \right), \qquad
	g_{ZHH_3^0} = i\sqrt{\frac{2}{3}}\frac{e}{s_Wc_W}\left(c_\alpha \frac{v_\phi}{v} - \sqrt{3} s_\alpha \frac{v_\chi}{v} \right), \nonumber \\
	g_{ZH_5^0 H_3^0} &=& -i \sqrt{\frac{1}{3}} \frac{e}{s_Wc_W}\frac{v_\phi}{v}, \qquad \qquad \qquad \qquad \quad
	g_{ZH_5^+H_3^{+*}} = \frac{e}{2 s_Wc_W}\frac{v_\phi}{v}, \nonumber \\
	g_{ZZh} &=& \frac{e^2}{2 s_W^2 c_W^2} \left(c_\alpha v_\phi - \frac{8}{\sqrt{3}} s_\alpha v_\chi \right),  \qquad \qquad
	g_{ZZH} = \frac{e^2}{2 s_W^2 c_W^2} \left(s_\alpha v_\phi + \frac{8}{\sqrt{3}} c_\alpha v_\chi \right), \nonumber \\
	g_{ZZ H_5^0} &=& -\sqrt{\frac{8}{3}} \frac{e^2}{s_W^2 c_W^2} v_{\chi}, \qquad \qquad \qquad \qquad \quad
	g_{Z W^+ H_5^{+*}} = -\frac{\sqrt{2} e^2}{c_W s_W^2} v_{\chi},
\end{eqnarray}
and the SM coupling $g_{ZZh}^{\rm SM}$ is given by
\begin{equation}
	g_{ZZh}^{\rm SM} = \frac{e^2 v}{2 s_W^2 c_W^2}.
\end{equation}
We use $s_{\alpha} \equiv \sin\alpha$, $c_{\alpha} \equiv \cos\alpha$, and similarly for the sine and cosine of the weak mixing angle.

For a reference SM Higgs mass of $m_h^{\rm SM} = 125$~GeV and setting $U = 0$, the global electroweak fit yields~\cite{Baak:2014ora}
\begin{equation}
	S_{\rm exp} = 0.06 \pm 0.09, \qquad \qquad
	T_{\rm exp} = 0.10 \pm 0.07,
\end{equation}
with a correlation $\rho_{ST} = +0.91$.  These values ({\tt MHREF}, {\tt SEXP}, {\tt DSEXP}, {\tt TEXP}, {\tt DTEXP}, and {\tt RHOST}, respectively) are hard-coded in the subroutine {\tt INITINDIR} in /src/gmindir.f.

We compute the $\chi^2$ according to
\begin{equation}
	\chi^2 =  \frac{1}{\left(1-\rho_{ST}^2\right)}\left[\frac{\left(S-S_{\rm exp}\right)^2}{\left( \Delta S_{\rm exp} \right)^2}+\frac{\left(T-T_{\rm exp}\right)^2}{\left( \Delta T_{\rm exp} \right)^2}-\frac{2 \rho_{ST} \left(S-S_{\rm exp}\right)\left(T-T_{\rm exp}\right)}{\Delta S_{\rm exp} \Delta T_{\rm exp}}\right],
\end{equation}
where $\Delta S_{\rm exp}$ and $\Delta T_{\rm exp}$ are the $1\sigma$ experimental uncertainties.

It is well known that the one-loop calculation of the $T$ parameter in the GM model yields a divergent result due to the explicit breaking of the custodial symmetry by hypercharge gauge interactions~\cite{Gunion:1990dt}.  In a proper treatment $T$ acquires a counterterm, which must be set, e.g., by specifying the energy scale at which the custodial symmetry in the scalar potential is exact.  Here we take the conservative approach of marginalizing over $T$, which amounts to setting 
\begin{equation}
	T = T_{\rm exp} + \rho_{ST} (S-S_{\rm exp})\frac{\Delta T_{\rm exp}}{\Delta S_{\rm exp}}.
\end{equation}
We set the flag ${\tt SPAROK} = 1$ if the GM prediction for the $S$ parameter yields $\chi^2 \leq 4$, and ${\tt SPAROK} = 0$ otherwise.

\subsection{$b \to s \gamma$}

The current world average experimental measurement of ${\rm BR}(\bar B \to X_s \gamma)$, for a photon energy $E_{\gamma} > 1.6$~GeV, is~\cite{Beringer:1900zz}
\begin{equation}
	{\rm BR}(\bar B \to X_s \gamma)_{\rm exp} = (3.55 \pm 0.24 \pm 0.09) \times 10^{-4}.
\end{equation}
To evaluate the constraint from this observable, we calculated the GM model predictions for a grid of $(m_3, v_{\chi})$ values by adapting the implementation for the Type-I 2HDM in the code SuperIso v3.3~\cite{SuperIso} (which makes use of the code 2HDMC v1.6.4~\cite{2HDMC}).  Our choice of input parameters yields a prediction in the limit $v_{\chi} \to 0$ or $m_3 \to \infty$ of
\begin{equation}
	{\rm BR}(\bar B \to X_s \gamma)_{\rm SM \, limit} = (3.11 \pm 0.23) \times 10^{-4},
\end{equation}
where the theoretical uncertainty is taken from Ref.~\cite{Misiak:2006zs}.  We scale the theoretical uncertainty by the ratio ${\rm BR}(\bar B \to X_s \gamma)_{\rm GM}/{\rm BR}(\bar B \to X_s \gamma)_{\rm SM \, limit}$ before combining it in quadrature with the experimental uncertainties.

The two data files /src/bsgtight.data and /src/bsgloose.data contain two sets of points $(m_3, v_{\chi})$ corresponding to the contour at which ${\rm BR}(\bar B \to X_s \gamma)_{\rm GM} = 2.88 \times 10^{-4}$ (``tight'' constraint) and $2.48 \times 10^{-4}$ (``loose'' constraint), respectively.  These correspond to a 2$\sigma$ deviation from the experimental central value (``tight'') and a value 2$\sigma$ ``worse'' than the SM prediction (``loose'').  For further explanation, see Ref.~\cite{indirect}.  Model points are checked for consistency with these constraints by linearly interpolating the upper bound on $v_{\chi}$ to the appropriate mass $m_3$.  For $m_3 < 10$~GeV the limit on $v_{\chi}$ for $m_3 = 10$~GeV is used, and for $m_3 > 1000$~GeV the limit on $v_{\chi}$ for $m_3 = 1000$~GeV is used.  (This latter limiting value falls outside the parameter range allowed by theoretical constraints, and so is irrelevant in practice.)

We set the flag {\tt BSGAMTIGHTOK} $=1$ if the GM prediction for ${\rm BR}(\bar B \to X_s \gamma)$ satisfies the ``tight'' 2$\sigma$ constraint, and {\tt BSGAMTIGHTOK} $=0$ otherwise.
Similarly, we set the flag {\tt BSGAMLOOSEOK} $=1$ if the GM prediction for ${\rm BR}(\bar B \to X_s \gamma)$ satisfies the ``loose'' 2$\sigma$ constraint, and {\tt BSGAMLOOSEOK} $=0$ otherwise.

\subsection{$B_s^0 \to \mu^+ \mu^-$}

The time-averaged branching ratio for $B_s^0 \to \mu^+ \mu^-$, normalized to its Standard Model value, is given to an excellent approximation by the ratio of $Z$-penguin contributions~\cite{indirect,Li:2014fea}
\begin{equation}
	{\tt RBSMM} \equiv 
	\frac{\overline {\rm BR}(B_s^0 \to \mu^+ \mu^-)}{\overline {\rm BR}(B_s^0 \to \mu^+ \mu^-)_{\rm SM}}
	\simeq \left| \frac{C_{10}^{\rm SM} + C_{10}^{\rm GM}}{C_{10}^{\rm SM}} \right|^2,
\end{equation}
where~\cite{Li:2014fea}
\begin{equation}
	C_{10}^{\rm SM} = -0.9380 \left[ \frac{M_t}{173.1~{\rm GeV}} \right]^{1.53}
	\left[ \frac{\alpha_s(M_Z)}{0.1184} \right]^{-0.09}
\end{equation}
and~\cite{indirect,Li:2014fea}
\begin{equation}
	C_{10}^{\rm GM} = C_{10}^{\rm SM} + \tan^2 \theta_H \frac{x_{tW}}{8}
	\left[ \frac{x_{t3}}{1-x_{t3}} + \frac{x_{t3} \log x_{t3}}{(1-x_{t3})^2} \right],
\end{equation}
with $x_{tW} = \overline m_{t}^2(M_t)/M_W^2$ and $x_{t3} = \overline m_t^2(M_t)/m_3^2$.\footnote{The calculation of the $\overline{\rm MS}$ running top quark mass $\overline m_t(\mu)$ is described in Sec.~\ref{sec:Hff}.  $M_t$ is the pole mass.}  For numerical stability we use an expansion in $\delta \equiv x_{t3}-1$ when $x_{t3} \simeq 1$ to within a part in $10^{-4}$, 
\begin{equation}
	\left[ \frac{x_{t3}}{1-x_{t3}} + \frac{x_{t3} \log x_{t3}}{(1-x_{t3})^2} \right] \simeq -\frac{1}{2} - \frac{\delta}{6}
	\qquad \qquad (\delta \equiv x_{t3} -1 \to 0).
\end{equation}

The corresponding SM prediction and its uncertainty are~\cite{Li:2014fea}
\begin{equation}
	\overline {\rm BR}(B_s^0 \to \mu^+ \mu^-)_{\rm SM} = (3.67 \pm 0.25) \times 10^{-9}
	\left|\left[ \frac{M_t}{173.1~{\rm GeV}} \right]^{1.53}
	\left[ \frac{\alpha_s(M_Z)}{0.1184} \right]^{-0.09} \right|^2.
\end{equation}
We calculate the prediction in the GM model by scaling this prediction and its uncertainty by {\tt RBSMM}.

The current world average experimental value (from CMS and LHCb) is~\cite{bsmmexp}
\begin{equation}
	\overline {\rm BR}(B_s^0 \to \mu^+\mu^-)_{\rm expt} = (2.9 \pm 0.7) \times 10^{-9}.
\end{equation}
The experimental central value ({\tt BMMEXP}) and its uncertainty ({\tt DBMMEXP}) are hard-coded in the subroutine {\tt INITINDIR} in /src/gmindir.f.

Combining the theoretical and experimental uncertainties in quadrature, this measured value is about $1\sigma$ below the SM prediction.  The GM prediction is always higher than the SM prediction (in worse agreement with experiment) and depends only on the parameters $m_3$ and $\tan\theta_H$.  

We set the flag ${\tt BSMMOK} = 1$ if the GM prediction for $\overline {\rm BR}(B_s^0 \to \mu^+\mu^-)$ is within $2\sigma$ of the experimental value, and ${\tt BSMMOK} = 0$ otherwise.

\section{Decays}
\label{sec:decays}

Starting from the tree-level masses and couplings, the code calculates the decay widths of the Higgs bosons into various final states.  At tree level the Higgs bosons can decay into pairs of fermions, pairs of massive gauge bosons, a gauge boson and a lighter Higgs boson, and two lighter Higgs bosons.  Decays of the neutral Higgs bosons into $gg$, $\gamma\gamma$, and $Z\gamma$ are induced at one loop.

\subsection{$H \to f \bar f^{\prime}$}
\label{sec:Hff}

The custodial singlet states $h$ and $H$ and the custodial triplet states $H_3^0$ and $H_3^{\pm}$ can decay to pairs of fermions.  The custodial fiveplet states do not couple to fermions.

The Feynman rule for a scalar coupling to $f \bar f^{\prime}$ is parameterized as $i (g^S + g^P \gamma_5)$, where $g^S$ is the scalar part and $g^P$ is the pseudoscalar part.  $g^S$ and $g^P$ can be simultaneously nonzero only for charged Higgs couplings to fermions.

The decay width to fermions is given by (the number of colors $N_c = 3$ for quarks and 1 for leptons)
\begin{equation}
	\Gamma(H \to f \bar f^{\prime}) = \frac{N_c m_H}{8 \pi}
	\left\{ \left[ 1 - (x_1 + x_2)^2 \right] |g^S|^2 + \left[ 1 - (x_1 - x_2)^2 \right] |g^P|^2 \right\}
	\lambda^{1/2}(x_1^2, x_2^2),
\end{equation}
where $x_1 = m_f/m_H$, $x_2 = m_{f^{\prime}}/m_H$, and the kinematic function $\lambda$ is given by
\begin{equation}
	\lambda(x,y) = (1 - x - y)^2 - 4xy.
\end{equation}

For scalar decays to quarks, we incorporate the QCD corrections as follows.
First, we incorporate the leading QCD corrections by replacing $m_q \to \overline m_q(M_H)$ in the Yukawa couplings $g^S$ and $g^P$, where $\overline m_q(M_H)$ is the $\overline{\rm MS}$ running quark mass evaluated at the scale of the parent Higgs particle's mass.  We compute the running quark masses using~\cite{Djouadi:1995gt}
\begin{equation}
	\overline m_q(\mu) = \overline m_q(M_q) \frac{c[\alpha_s(\mu)/\pi]}{c[\alpha_s(\mu)/\pi]},
\end{equation}
where
\begin{eqnarray}
	c(x) &=& \left( \frac{25}{6}x \right)^{12/25} (1 + 1.014 x + 1.389 x^2), 
		\qquad M_c < \mu < M_b \nonumber \\
	c(x) &=& \left( \frac{23}{6} x \right)^{12/23} (1 + 1.175 x + 1.501 x^2),
		\qquad M_b < \mu.
\end{eqnarray}
The running strong coupling constant is computed using~\cite{Djouadi:1995gt}
\begin{equation}
	\alpha_s^{(N_f)}(\mu) = \frac{12 \pi}{(33 - 2N_f) \log(\mu^2/\Lambda_{N_f}^2)}
		\left[ 1 - 6 \frac{(153 - 19N_f)}{(33 - 2N_f)^2} 
		\frac{\log \log(\mu^2/\Lambda_{N_f}^2)}{\log(\mu^2/\Lambda_{N_f}^2)} \right].
\end{equation}
We implement matching at the bottom quark threshold by requiring continuity of $\alpha_s$.
Above the top threshold we continue to use the five-flavor scheme for consistency with HDECAY~\cite{Djouadi:1997yw}.

Second, for decays of neutral CP-even scalars to $b \bar b$ or $c \bar c$ we incorporate the finite QCD corrections by multiplying the partial width given above by the factor~\cite{Djouadi:1995gt}
\begin{equation}
	\left[ \Delta_{QCD} + \Delta_t \right],
\end{equation}
where
\begin{eqnarray}
	\Delta_{QCD} &=& 1 + 5.67 \frac{\alpha_s(M_H)}{\pi}
		+ (35.94 - 1.36 N_f) \left( \frac{\alpha_s(M_H)}{\pi} \right)^2, \nonumber \\
	\Delta_t &=& \left( \frac{\alpha_s(M_H)}{\pi} \right)^2 
		\left[ 1.57 - \frac{2}{3} \log(M_H^2/M_t^2) 
		+ \frac{1}{9} \log^2 (\overline m_q^2(M_H)/M_H^2) \right].
\end{eqnarray}

The relevant SM inputs to GMCALC are 
\begin{equation}
	{\tt ALSMZ} = \alpha_s(M_Z), \qquad {\tt MTPOLE} = M_t, \qquad
	{\tt MBMB} = \overline m_b(m_b), \qquad {\tt MCMC} = \overline m_c(m_c).
\end{equation}
The values are set in {\tt INITIALIZE\_SM} in /src/gminit.f.
The $b$ and $c$ quark pole masses, and the running top quark mass, are obtained using the $\mathcal{O}(\alpha_s)$ relation~\cite{Djouadi:1995gt}
\begin{equation}
	\overline m_q(M_q) = M_q / [1 + 4 \alpha_s / 3 \pi].
\end{equation}

\subsection{$H \to V_1 V_2$}

The custodial singlet states $h$ and $H$, as well as the neutral custodial fiveplet state $H_5^0$, can decay to $W^+W^-$ and $ZZ$.  The charged custodial fiveplet state $H_5^+$ can decay to $W^+Z$.  The doubly-charged custodial fiveplet state $H_5^{++}$ can decay to $W^+W^+$.  The custodial triplet states do not couple to pairs of massive vector bosons.

The Feynman rule for a scalar coupling to massive vector bosons $V_1^{\mu} V_2^{\nu}$ is parameterized as $i g_{HV_1V_2} g^{\mu\nu}$.

The on-shell two-body decay width into two massive vector bosons is given by
\begin{equation}
	\Gamma(H \to V_1 V_2) 
	= S_V \frac{|g_{HV_1V_2}|^2 m_H^3}{64 \pi M_{V_1}^2 M_{V_2}^2}
	\left[ 1 - 2 k_1 - 2 k_2 + 10 k_1 k_2 + k_1^2 + k_2^2 \right]
	\lambda^{1/2}(k_1, k_2),
\end{equation}
where $k_1 = M_{V_1}^2/m_H^2$ and $k_2 = M_{V_2}^2/m_H^2$, and $S_V$ is a symmetry factor given by $S_V = 1$ if $V_1$ and $V_2$ are distinct bosons (e.g., $W^+W^-$ or $Z W^+$) and $S_V = 1/2$ if $V_1$ and $V_2$ are identical bosons (e.g., $ZZ$ or $W^+W^+$).

We also implement decays of $h$, $H$, $H_5^0$, and $H_5^{++}$ to $WW^*$ and $ZZ^*$ (with one of the two gauge bosons off-shell) when the scalar mass is below threshold for the on-shell two-body decay.  We have not yet implemented the singly off-shell decay $\Gamma(H_5^+ \to W^+ Z^* + W^{+*} Z)$.  Following Ref.~\cite{Djouadi:1995gv}, for decays to two vector bosons with the same mass $M_V$,
\begin{eqnarray}
	\Gamma(H \to V V^*) &=& S_V \delta_V \frac{3 |g_{HVV}|^2 m_H}{64 \pi^3 v^2} 
	\left[ \frac{1 - 8 k + 20 k^2}{(4 k - 1)^{1/2}} 
	\arccos \left( \frac{3 k - 1}{2 k^{3/2}} \right)
	- \frac{1-k}{6k} (2 - 13 k + 47 k^2) \right. \nonumber \\
	&& \left. \qquad \qquad \qquad \qquad
	- \frac{1}{2} (1 - 6k + 4 k^2) \log k \right],
	\label{eq:GamHVV*}
\end{eqnarray}
where $k = M_V^2/m_H^2$ and the factors $\delta_W$ and $\delta_Z$ are\footnote{We absorb the factor of $c_W^4$ that appears in the denominator of $\delta_Z$ in Eq.~(36) of Ref.~\cite{Djouadi:1995gv} into the coupling $|g_{HZZ}|^2$.  We also separate out a symmetry factor of $2$ from $\delta_W$ for later convenience when implementing singly-offshell $H_1 \to V^* H_2$ decays.}
\begin{eqnarray}
	\delta_W &=& \frac{3}{2}, \nonumber \\
	\delta_Z &=& 3 \left( \frac{7}{12} - \frac{10}{9} s_W^2 + \frac{40}{27} s_W^4 \right).
	\label{eq:dwdz}
\end{eqnarray}
We have not taken into account the interference effects in same-flavor decays due to crossed diagrams.

\subsection{$H_1 \to V H_2$}

The custodial singlet states $h$ and $H$ can decay to a vector boson plus a custodial triplet scalar.  The custodial triplet states $H_3^0$ and $H_3^{\pm}$ can decay to a vector boson plus a custodial singlet state, or to a vector boson plus a custodial fiveplet state.  The custodial fiveplet states $H_5^0$, $H_5^{\pm}$, and $H_5^{\pm \pm}$ can decay to a vector boson plus a custodial triplet state.  

The Feynman rule for the $H_1 H_2^* V^*_{\mu}$ coupling (all particles and momenta incoming) is parameterized as $i g_{V^*H_1H_2^*} (p_1 - p_2)_{\mu}$, where $p_1$ ($p_2$) is the incoming momentum of the scalar $H_1$ ($H_2^*$).

The on-shell two-body decay width into one vector and one lighter scalar is given by
\begin{equation}
	\Gamma(H_1 \to V H_2) = \frac{|g_{V^*H_1H_2^*}|^2 M_V^2}{16 \pi m_{H_1}}
	\lambda\left( \frac{m_{H_1}^2}{M_V^2}, \frac{m_{H_2}^2}{M_V^2} \right)
	\lambda^{1/2} \left( \frac{M_V^2}{m_{H_1}^2}, \frac{m_{H_2}^2}{m_{H_1}^2} \right).
\end{equation}
Here $V$ denotes one of the gauge bosons $Z$, $W^+$, or $W^-$, such that the decays $H^0 \to W^+ H^-$ and $H^0 \to W^- H^+$ are distinct.

We also implement $H_1 \to V^* H_2$ decays (with the gauge boson off-shell) when the $H_1$ mass is below threshold for the on-shell two-body decay.  Following Ref.~\cite{Djouadi:1995gv},
\begin{equation}
	\Gamma(H_1 \to V^* H_2) = \delta_V \frac{3 |g_{V^* H_1 H_2^*} |^2 M_V^2 m_{H_1}}
	{16 \pi^3 v^2} G_{H_2 V},
\end{equation}
where again $V$ denotes one of the gauge bosons $Z$, $W^+$, or $W^-$, such that the decays $H^0 \to W^+ H^-$ and $H^0 \to W^- H^+$ are distinct.  $\delta_W$ and $\delta_Z$ were given in Eq.~(\ref{eq:dwdz}).  The kinematic function $G_{ij}$ is defined as follows (here we fix a typing error in Ref.~\cite{Djouadi:1995gv} as pointed out in Ref.~\cite{Akeroyd:1998dt}: the last term is $+2 \lambda_{ij}/k_j$ rather than $-2 \lambda_{ij}/k_j$):
\begin{eqnarray}
	G_{ij} &=& \frac{1}{4} \left\{ 2 (-1 + k_j - k_i) \sqrt{\lambda_{ij}} 
	\left[ \frac{\pi}{2} + \arctan \left( \frac{k_j (1 - k_j + k_i) - \lambda_{ij}}{(1 - k_i) \sqrt{\lambda_{ij}}} \right) \right] \right. \nonumber \\
	&& \qquad \left. + (\lambda_{ij} - 2 k_i) \log k_i 
	+ \frac{1}{3} (1 - k_i) \left[ 5 (1 + k_i) - 4 k_j + \frac{2 \lambda_{ij}}{k_j} \right] \right\},
\end{eqnarray}
where $k_i \equiv k_{H_2} = m_{H_2}^2/m_{H_1}^2$, $k_j \equiv k_V = M_V^2/m_{H_1}^2$, and 
\begin{equation}
	\lambda_{ij} = -1 + 2 k_i + 2 k_j - (k_i - k_j)^2.
\end{equation}

\subsection{$H_1 \to H_2 H_3$}

The custodial singlet states $h$ and $H$ can decay into a pair of custodial triplet states or a pair of custodial fiveplet states.  Furthermore $H$ can decay into $hh$.  The custodial fiveplet states $H_5^0$, $H_5^{\pm}$, and $H_5^{\pm \pm}$ can decay into a pair of custodial triplet states.  The custodial triplet states cannot decay into pairs of scalars due to a combination of custodial SU(2) invariance and Bose symmetry.

The Feynman rule for the $H_1 H_2^* H_3^*$ coupling (all particles incoming) is parameterized as $-i g_{123}$.

The decay width for $H_1$ into two lighter scalars $H_2 H_3$ is
\begin{equation}
	\Gamma(H_1 \to H_2 H_3) = S_H \frac{|g_{123}|^2}{16 \pi m_{H_1}}
	\lambda^{1/2}(X_2, X_3),
\end{equation}
where $X_2 = m_{H_2}^2/m_{H_1}^2$ and $X_3 = m_{H_3}^2/m_{H_1}^2$, and $S_H$ is a symmetry factor given by $S_H = 1$ if $H_2$ and $H_3$ are distinct bosons and $S_H = 1/2$ if $H_2$ and $H_3$ are identical bosons.

\subsection{$H \to \gamma\gamma$}

Neutral scalar decays into two photons proceed through a loop of charged particles.  The width is given by~\cite{HHG}
\begin{equation}
	\Gamma(H \to \gamma\gamma) = \frac{\alpha_{EM}^2 m_H^3}{256 \pi^3 v^2}
	| \mathcal{A}_H^{\gamma\gamma} |^2,
\end{equation}
where $\alpha_{EM}$ is the electromagnetic fine-structure constant, $v = (\sqrt{2} G_F)^{-1/2} \simeq 246$~GeV is the SM Higgs vacuum expectation value, and $\mathcal{A}_H^{\gamma\gamma}$ represents the sum of the loop amplitudes for initial particle $H$.

For an initial scalar ($S = h$, $H$, or $H_5^0$), the amplitude receives contributions from fermions, $W$ bosons, and charged Higgs bosons ($H_3^+$, $H_5^+$, and $H_5^{++}$) in the loop, and is given by
\begin{equation}
	\mathcal{A}_S^{\gamma\gamma} 
	= \kappa_f^S \sum_f N_{cf} Q_f^2 F_{1/2}(\tau_f)
	+ \kappa_W^S F_1(\tau_W)
	+ \sum_s \beta_s^S Q_s^2 F_0(\tau_s).
\end{equation}
For the fermion loops, $N_{cf}$ and $Q_f$ are the number of colors and electric charge in units of $e$, respectively, for fermion $f$, and $\kappa_f^S$ is the scaling factor for the coupling of $S$ to fermions relative to the corresponding coupling of the SM Higgs boson, defined in such a way that the Feynman rule for the $S f \bar f$ coupling is $-i (m_f/v) \kappa^S_f$.  The custodial fiveplet does not couple to fermions, so $\kappa_f^{H_5^0} = 0$.

For the $W$ loop, $\kappa_W^S$ is the scaling factor for the coupling of $S$ to $W$ pairs relative to the corresponding coupling of the SM Higgs boson, defined so that the $S W^+_{\mu} W^-_{\nu}$ Feynman rule is $i \kappa_W^S (2 M_W^2/v) g_{\mu\nu}$.

For the scalar loops, the sum over $s$ runs over all electrically charged scalars in the GM model ($H_3^+$, $H_5^+$, and $H_5^{++}$).  $Q_s$ is the electric charge of scalar $s$ in units of $e$, and $\beta_s^S = g_{Sss^*} v/2 m_s^2$.  The coupling $g_{Sss^*}$ is defined in such a way that the corresponding interaction Lagrangian term is $\mathcal{L} \supset - g_{Sss^*} S s s^*$.

The loop factors are given in terms of the usual functions~\cite{HHG}, 
\begin{eqnarray}
	F_1(\tau) &=& 2+3\tau+3\tau(2-\tau) f(\tau), \nonumber \\
	F_{1/2}(\tau) &=& -2\tau[1+(1-\tau) f(\tau)], \nonumber \\
	F_0(\tau) &=& \tau[1-\tau f(\tau)],
\end{eqnarray}
where
\begin{equation}
	f(\tau) = \left\{ \begin{array}{l l}
	\left[\sin^{-1} \left(\sqrt{\frac{1}{\tau}}\right) \right]^2 & \quad  {\rm if} \ \tau \geq 1, \\
	-\frac{1}{4}\left[ \log \left(\frac{\eta_+}{\eta_-}\right) - i \pi \right]^2 & \quad  {\rm if} \ \tau < 1, \\
	\end{array} \right.
	\label{feq}
\end{equation}
with $\eta_{\pm} = 1 \pm \sqrt{1-\tau}$.  The argument is $\tau_i \equiv 4 m_i^2/ m_h^2$. 

For an initial pseudoscalar ($A = H_3^0$), the amplitude receives contributions only from fermions in the loop, and is given by
\begin{equation}
	\mathcal{A}_A^{\gamma\gamma} = \kappa_f^A \sum_f N_{cf} Q_f^2 F_{1/2}^A(\tau_f)
\end{equation}
where the Feynman rule for the $A f \bar f$ coupling is defined as $- (m_f/v) \kappa_f^A \gamma_5$ and the loop function is
\begin{equation}
	F_{1/2}^A(\tau) = -2 \tau f(\tau).
\end{equation}

\subsection{$H \to gg$}

Neutral scalar decays to two gluons proceed through a loop of colored particles.  In the GM model, the only colored particles are the SM quarks.  Therefore this decay occurs only for $h$, $H$, and $H_3^0$ (the custodial fiveplet does not couple to fermions).

The width is given by~\cite{HHG}
\begin{equation}
	\Gamma(H \to gg) = \frac{\alpha_s^2 m_H^3}{128 \pi^3 v^2} |\mathcal{A}_H^{gg}|^2,
\end{equation}
where $\mathcal{A}_H^{gg}$ represents the sum of the loop amplitudes for initial particle $H$.

For an initial scalar ($S = h$ or $H$), the amplitude is
\begin{equation}
	\mathcal{A}_S^{gg} = \kappa_f^S \sum_f F_{1/2}(\tau_f).
\end{equation}

For an initial pseudoscalar ($A = H_3^0$), the amplitude is
\begin{equation}
	\mathcal{A}_A^{gg} = \kappa_f^A \sum_f F_{1/2}^A(\tau_f).
\end{equation}

We incorporate the QCD corrections as follows.  First, we evaluate $\alpha_s$ in the leading-order amplitude at the scale of the parent particle's mass.  Second, for the decays of CP-even neutral scalars, we multiply the leading order amplitude by the factor~\cite{Djouadi:1995gt}
\begin{equation}
	\left[ 1 + E^{N_f} \alpha_s^{(N_f)}/\pi \right],
\end{equation}
where
\begin{equation}
	E^{N_f} = \frac{95}{4} - \frac{7}{6} N_f + \frac{33 - 2N_f}{6} \log(\mu^2/M_H^2),
\end{equation}
and we use $N_f = 5$ throughout, consistent with {\tt NF-GG} $= 5$ in HDECAY~\cite{Djouadi:1997yw}.

\subsection{$H \to Z \gamma$}

Neutral scalar decays to $Z$ plus a photon proceed through a loop of charged particles.  The width is given by~\cite{HHG}
\begin{equation}
	\Gamma(H \to Z \gamma) = \frac{\alpha_{EM}^2 m_H^3}{128 \pi^3 v^2}
	| \mathcal{A}_H^{Z\gamma} |^2 \left( 1 - \frac{M_Z^2}{m_H^2} \right)^3,
\end{equation}
where $\mathcal{A}_H^{Z\gamma}$ represents the sum of the loop amplitudes for initial particle $H$.

For an initial scalar ($S = h$, $H$, or $H_5^0$), the amplitude is\footnote{Decays of $H_5^0 \to Z\gamma$ may also receive a contribution from mixed loops containing both $H_5^+$ and $W^+$.  These contributions have not been implemented yet.}
\begin{equation}
	\mathcal{A}_S^{Z\gamma} = \kappa_f^S A_f + \kappa_V^S A_W + \frac{v}{2} A_s,
\end{equation}
where the contributions from fermions, $W$ bosons, and scalars are given by~\cite{HHG}
\begin{eqnarray}
	A_f &=& \sum_f N_{cf} 
	\frac{-2 Q_f \left(T^{3L}_f - 2 Q_f \sin^2\theta_W\right)}{\sin\theta_W\cos\theta_W}
	\left[ I_1(\tau_f,\lambda_f)-I_2(\tau_f,\lambda_f) \right], \nonumber \\
	A_W &=& -\cot\theta_W\left\{4\left(3-\tan^2\theta_W\right) I_2\left(\tau_W,\lambda_W\right)+\left[\left(1+\frac{2}{\tau_W}\right)\tan^2\theta_W-\left(5+\frac{2}{\tau_W}\right)\right] I_1\left(\tau_W,\lambda_W\right)\right\}, \nonumber \\
	A_s &=& \sum_s 2 \frac{g_{hss^*}\,C_{Zss^*}\,Q_s}{m_s^2}
	I_1\left( \tau_s, \lambda_s \right).
	\label{eq:Zgaamps}
\end{eqnarray}
Here $T^{3L}_f = \pm 1/2$ is the third component of isospin for the left-handed fermion $f$.  The scalar amplitude depends on the coupling $C_{Zss^*} \equiv g_{Zss^*}/e$ of the scalar to the $Z$ boson, defined in  such a way that the corresponding coupling of the scalar to the photon is $C_{\gamma s s^*} \equiv g_{\gamma s s^*}/e = Q_s$.

The loop factors are given in terms of the functions~\cite{HHG}
\begin{eqnarray}
	 I_1(a,b) &=& \frac{ab}{2(a-b)} + \frac{a^2b^2}{2(a-b)^2} \left[f(a) - f(b)\right]
	 + \frac{a^2b}{(a-b)^2} \left[g(a) - g(b)\right], \nonumber \\
	 I_2(a,b) &=& -\frac{ab}{2(a-b)} \left[f(a) - f(b)\right],
\end{eqnarray}
where the function $f(\tau)$ was given in Eq.~(\ref{feq}) and
\begin{equation}
	g(\tau) = \left\{ \begin{array}{l l}
	\sqrt{\tau-1} \sin^{-1} \left(\sqrt{\frac{1}{\tau}}\right) & \quad  {\rm if} \ \tau \geq 1, \\
	\frac{1}{2} \sqrt{1-\tau} \left[ \log \left(\frac{\eta_+}{\eta_-}\right) - i \pi \right] 
		& \quad  {\rm if} \ \tau < 1,
	\end{array} \right.
	\label{geq}
\end{equation}
with $\eta_{\pm}$ defined as for $f(\tau)$.  The arguments of the functions are $\tau_i \equiv 4 m_i^2/m_h^2$ as before and $\lambda_i \equiv 4 m_i^2/M_Z^2$.

For an initial pseudoscalar ($A = H_3^0$), the amplitude is
\begin{equation}
	\mathcal{A}_A^{Z\gamma} = \kappa_f^A \sum_f N_{cf} 
	\frac{-2 Q_f \left(T^{3L}_f - 2 Q_f \sin^2\theta_W\right)}{\sin\theta_W\cos\theta_W}
	\left[ -I_2(\tau_f,\lambda_f) \right].
\end{equation}



\section{Using the GMCALC program}
\label{sec:using}

The GMCALC code package is available for download as a .tar.gz file from the web page
\begin{quote}
	http://people.physics.carleton.ca/$\sim$logan/gmcalc/
\end{quote}
The package includes this manual.  Feature requests and bug reports should be sent to Heather Logan at logan@physics.carleton.ca .

\subsection{Sample main programs provided with the code}

Three sample main programs are provided with the code.  These can be used as-is, or as templates for the user to write their own programs.  The command
\begin{quotation}
	{\tt \$ make sample}
\end{quotation}
compiles the program sample.f into an executable sample.x using gfortran.  The executable is run using
\begin{quotation}
	{\tt \$ ./sample.x}
\end{quotation}
The sample programs are as follows:
\begin{itemize}
\item {\tt gmpoint.f} performs the full set of available calculations for a single parameter point and outputs the spectrum, couplings, and decay tables to the terminal.
\item {\tt gmscan.f} performs a scan over the allowed parameter ranges using the approach described in Sec.~\ref{sec:scans}.  For each scan point allowed by theoretical and indirect experimental constraints, it writes a selection of observables to a file scan\_output.data.
\item {\tt gmmg5.f} generates a param\_card.dat file for use with the FeynRules model implementation.
\end{itemize}

\subsection{Setting the model parameters}

There are currently three choices of input parameters implemented in GMCALC:
\begin{itemize}

\item {\tt INPUTSET} $= 1$ uses the primary inputs $\mu_3^2$, $\lambda_1$, $\lambda_2$, $\lambda_3$, $\lambda_4$, $\lambda_5$, $M_1$, and $M_2$.  The parameter $\mu_2^2$ is set using the constraint on $v_{\phi}^2 + 8 v_{\chi}^2$ in terms of $G_F$.

\item {\tt INPUTSET} $= 2$ uses the primary inputs $\mu_3^2$, $m_h$, $\lambda_2$, $\lambda_3$, $\lambda_4$, $\lambda_5$, $M_1$, and $M_2$.  The parameter $\mu_2^2$ is again set using the constraint on $v_{\phi}^2 + 8 v_{\chi}^2$ in terms of $G_F$.

\item {\tt INPUTSET} $= 3$ uses the primary inputs $m_h$, $m_H$, $m_3$, $m_5$, $\sin\theta_H$, $\sin\alpha$, $M_1$, and $M_2$.  $G_F$ is also used to set $\mu_2^2$.
\end{itemize}
These inputs can be hand-coded in the sample programs (indicated by {\tt INPUTMODE} $= 0$).  Alternatively, the program can be run in interactive mode ({\tt INPUTMODE} $= 1$) in which case the user will be prompted to enter the inputs at the terminal.  In either case, the subroutine {\tt LOAD\_INPUTS} processes the inputs and computes the remaining potential parameters.  {\tt LOAD\_INPUTS} sets a flag {\tt INPUTOK} $= 1$ if the specified inputs yield an acceptable scalar potential.

\subsection{Checking consistency and computing the spectrum}

Before computing the physical spectrum, the scalar potential should be checked for consistency with theoretical constraints.  This is accomplished by the subroutine {\tt THYCHECK}, which returns three flags: {\tt UNIOK} $= 1$ indicates that the perturbative unitarity constraints on $\lambda_{1-5}$ are satisfied; {\tt BFBOK} $= 1$ indicates that the scalar potential is bounded from below; and {\tt MINOK} $= 1$ indicates that the desired electroweak-breaking vacuum is the global minimum of the potential.

The physical masses, vevs, and custodial-singlet mixing angle $\alpha$ can then be computed by the subroutine {\tt CALCPHYS}.  Results are passed via the common block
\begin{quote}
	{\tt COMMON/PHYSPARAMS/MHL,MHH,MH3,MH5,ALPHA,VPHI,VCHI}.
\end{quote}
They can be accessed directly by adding this common block declaration in one of the sample programs; alternatively, they can be output to the terminal by the subroutine {\tt PRINT\_RESULTS} (see Sec.~\ref{sec:output}).

With the physical spectrum computed, the indirect constraints can be checked by calling the subroutine {\tt CALCINDIR}.  This returns a series of flags which, if set to $1$, indicate that the model point satisfies the corresponding indirect constraint.  The flags are: {\tt BSMMOK} ($B_s^0 \to \mu^+\mu^-$), {\tt SPAROK} (oblique $S$ parameter), {\tt BSGAMLOOSEOK} (``loose'' constraint on $b \to s \gamma$), and {\tt BSGAMTIGHTOK} (``tight'' constraint on $b \to s \gamma$).  These can be accessed directly by including the common block
\begin{quote}
      {\tt COMMON/INDIR/RBSMM,SPARAM,BSMMOK,SPAROK,BSGAMLOOSEOK,BSGAMTIGHTOK}.
\end{quote}
They are also output to the terminal by the subroutine {\tt PRINT\_RESULTS} (see Sec.~\ref{sec:output}).  The double precision variables {\tt RBSMM} and {\tt SPARAM} in this common block contain the ratio of $\overline {\rm BR}(B_s^0 \to \mu^+\mu^-)$ to its SM value and the value of the $S$ parameter for this model point, respectively.

\subsection{Computing couplings and decays}

Once {\tt CALCPHYS} has been called, we are ready to compute Higgs couplings and/or decay branching ratios.  There are three subroutines that can be called independently of each other:
\begin{itemize}
\item {\tt HLCOUPS} computes the kappa factors $\kappa_i^h$ (i.e., the couplings normalized to their SM values) of $h$.  These are output to the terminal in a tidy form by {\tt PRINT\_HCOUPS}, but can also be accessed through the common block
\begin{quote}
	{\tt COMMON/KAPPASL/KVL,KFL,KGAML,KZGAML,DKGAML,DKZGAML}.
\end{quote}

\item {\tt HHCOUPS} does the same but for $H$.  These are output to the terminal by {\tt PRINT\_HCOUPS}, but can also be accessed through the common block 
\begin{quote}
	{\tt COMMON/KAPPASH/KVH,KFH,KGAMH,KZGAMH,DKGAMH,DKZGAMH}.
\end{quote}

\item {\tt CALCDECAYS} performs the full set of partial width calculations (see Sec.~\ref{sec:decays}) for all the scalar particles in the model, as well as for the top quark, which can decay to $H_3^+ b$ if kinematically allowed.  The resulting branching ratios and total widths are output to the terminal in a tidy form by {\tt PRINT\_DECAYS}, but can also be accessed through the series of common blocks for each particle as follows:
\begin{quote}
	$h$: {\tt COMMON/HLBRS/HLBRB, HLBRTA, HLBRMU, HLBRS, HLBRC, HLBRT,
      HLBRG, HLBRGA, HLBRZGA, HLBRW, HLBRZ, 
          HLBRWH3P, HLBRZH3N,
          HLBRH3N, HLBRH3P, HLBRH5N, HLBRH5P, HLBRH5PP, HLWDTH} \\
     $H$: {\tt COMMON/HHBRS/HHBRB, HHBRTA, HHBRMU, HHBRS, HHBRC, HHBRT,
          HHBRG, HHBRGA, HHBRZGA, HHBRW, HHBRZ, 
          HHBRWH3P, HHBRZH3N,
          HHBRHL, HHBRH3N, HHBRH3P, HHBRH5N, HHBRH5P, HHBRH5PP, 
          HHWDTH} \\
          $H_3^0$: {\tt COMMON/H3NBRS/H3NBRB, H3NBRTA, H3NBRMU, H3NBRS, H3NBRC, H3NBRT,
         H3NBRZHL, H3NBRZHH, H3NBRZH5N, H3NBRWH5P,
          H3NBRG, H3NBRGA, H3NBRZGA,
          H3NWDTH} \\
          $H_3^+$: {\tt COMMON/H3PBRS/H3PBRBC, H3PBRTA, H3PBRMU, H3PBRSU,
         H3PBRCS, H3PBRTB, H3PBRBU,
          H3PBRWHL, H3PBRWHH, H3PBRZH5P, H3PBRWH5N, H3PBRWH5PP,
          H3PWDTH} \\
          $H_5^0$: {\tt COMMON/H5NBRS/H5NBRGA, H5NBRZGA, H5NBRW, H5NBRZ,
          H5NBRZH3N, H5NBRWH3P, 
          H5NBRH3N, H5NBRH3P,
          H5NWDTH} \\
	$H_5^+$: {\tt COMMON/H5PBRS/H5PBRWZ, H5PBRZH3P, H5PBRWH3N, H5PBRH3PN, H5PWDTH} \\
	$H_5^{++}$: {\tt COMMON/H5PPBRS/H5PPBRWW, H5PPBRWH3, H5PPBRH3P, H5PPWDTH} \\
	$t$: {\tt COMMON/TOPBRS/TOPBRW, TOPBRH3P, TOPWDTH}
\end{quote}
\end{itemize}

\subsection{Outputs}
\label{sec:output}

There are three subroutines dedicated to printing results to the terminal:
\begin{itemize}
\item {\tt PRINT\_RESULTS} prints the Lagrangian parameters, the flags indicating theoretical consistency and consistency with indirect experimental constraints, and the physical masses, vevs, and custodial-singlet mixing angle.  These must have been previously computed by calls to {\tt LOAD\_INPUTS}, {\tt THYCHECK}, {\tt CALCPHYS} and {\tt CALCINDIR} (in that order).
\item {\tt PRINT\_HCOUPS} prints the kappa factors for $h$ and $H$.  These must have been previously computed by calls to the subroutines {\tt HLCOUPS} and {\tt HHCOUPS}.  
\item {\tt PRINT\_DECAYS} prints out the decay branching ratios and total widths of all the scalars in the model, as well as those of the top quark.  These must have been previously computed by a call to the subroutine {\tt CALCDECAYS}.
\end{itemize}

\subsection{Parameter scans}
\label{sec:scans}

To perform scans over the model parameters in an efficient way, the following strategy can be adopted.  Setting $m_h$ equal to the observed Higgs boson mass $\sim 125$~GeV and setting $\mu_2^2$ using $G_F$, the seven free parameters are ({\tt INPUTSET = 2})
\begin{equation}
	\mu_3^2, \lambda_2, \lambda_3, \lambda_4, \lambda_5, M_1, \ {\rm and} \ M_2.
\end{equation}

The parameters $\lambda_3$ and $\lambda_4$ are mainly constrained by the unitarity and bounded-from-below conditions.  The allowed range of $\lambda_3$ is
\begin{equation}
	- \frac{1}{2} \pi < \lambda_3 < \frac{3}{5} \pi.
\end{equation}
The allowed range of $\lambda_4$ is then
\begin{eqnarray}
	{\rm For} \ \lambda_3 < 0: && 
		-\lambda_3 < \lambda_4 < \left( -\frac{7}{11}\lambda_3 + \frac{2}{11} \pi \right), 
		\nonumber \\ 
	{\rm For} \ \lambda_3 \geq 0: && 
		-\frac{1}{3} \lambda_3 < \lambda_4 < \left( -\frac{7}{11}\lambda_3 + \frac{2}{11} \pi \right).
\end{eqnarray}

The parameter $\lambda_2$ is constrained by the first of the unitarity constraints in Eq.~(\ref{eq:uni}).  Since we don't know $\lambda_1$ until the rest of the parameters are set, we allow it to vary to obtain the least stringent constraint (which occurs when $\lambda_1 = 0$),
\begin{equation}
	|\lambda_2| < \frac{1}{3} \sqrt{4 \pi^2 - 2 \pi (7 \lambda_3 + 11 \lambda_4)}.
\end{equation}
Note that $0 < (7 \lambda_3 + 11 \lambda_4) < 2 \pi$.  Implementing a lower bound on the scan range for $\lambda_2$ from the bounded-from-below constraint does not dramatically improve the code's efficiency.

The last of the unitarity constraints in Eq.~(\ref{eq:uni}) then constrains 
\begin{equation}
	(-2 \pi + \lambda_2) < \lambda_5 < (2 \pi + \lambda_2).
\end{equation}

The dimensionful parameters $\mu_3^2$, $M_1$, and $M_2$ are constrained by the requirement that there be an acceptable electroweak symmetry breaking vacuum.  We find that the following ranges capture all allowed parameter points:
\begin{eqnarray}
	\mu_3^2 &>& -(200~{\rm GeV})^2, \nonumber \\
	M_1 &<& {\rm max}\left(3500~{\rm GeV}, 3.5 \sqrt{|\mu_3^2|}\right), \nonumber \\
	|M_2| &<& {\rm max}\left(250~{\rm GeV}, 1.3 \sqrt{|\mu_3^2|}\right).
\end{eqnarray}
Note that $M_1$ can be chosen positive with no loss of generality, so that $0 \leq M_1$.  $M_2$ takes either sign.  There is no upper bound on $\mu_3^2$; the limit $\mu_3^2 \gg v^2$ is the decoupling limit, in which the masses-squared of the predominantly-triplet states approach $\mu_3^2$.

\subsection{Standard Model inputs}

The Standard Model input parameters are initialized by the subroutine {\tt INITIALIZE\_SM}, which must be called before anything else.  The parameter values are hard-coded in /src/gminit.f.

The SM Higgs vev is computed as $v = (\sqrt{2} G_F)^{-1/2}$ and $M_W$ is computed using the tree-level relationship from $G_F$, $M_Z$, and $\alpha_{EM}$.

\section*{Acknowledgments}
This work was supported by the Natural Sciences and Engineering Research 
Council of Canada.  K.H.\ was also supported by the Government of Ontario through an Ontario Graduate Scholarship.





\begin{thebibliography}{99}

\bibitem{Georgi:1985nv} 
  H.~Georgi and M.~Machacek,
  Nucl.\ Phys.\ B {\bf 262}, 463 (1985).
  
\bibitem{Chanowitz:1985ug} 
  M.~S.~Chanowitz and M.~Golden,
  Phys.\ Lett.\ B {\bf 165}, 105 (1985).

\bibitem{HKL}
  K.~Hartling, K.~Kumar and H.~E.~Logan,
  Phys.\ Rev.\ D {\bf 90}, 015007 (2014)
  [arXiv:1404.2640 [hep-ph]].

\bibitem{indirect}
  K.~Hartling, K.~Kumar and H.~E.~Logan,
  arXiv:1410.5538 [hep-ph].

\bibitem{Aoki:2007ah} 
  M.~Aoki and S.~Kanemura,
  Phys.\ Rev.\ D {\bf 77}, 095009 (2008)
  [arXiv:0712.4053 [hep-ph]];   
  erratum Phys.\ Rev.\ D {\bf 89}, 059902 (2014).
  
\bibitem{Baak:2014ora} 
  M.~Baak, J.~Cuth, J.~Haller, A.~Hoecker, R.~Kogler, K.~Moenig, M.~Schott and J.~Stelzer,
  arXiv:1407.3792 [hep-ph].

\bibitem{Gunion:1990dt} 
  J.~F.~Gunion, R.~Vega and J.~Wudka,
  Phys.\ Rev.\ D {\bf 43}, 2322 (1991).
  
\bibitem{Beringer:1900zz} 
  J.~Beringer {\it et al.}  [Particle Data Group Collaboration],
  Phys.\ Rev.\ D {\bf 86}, 010001 (2012).
  
\bibitem{SuperIso}
F.~Mahmoudi,
  Comput.\ Phys.\ Commun.\  {\bf 178}, 745 (2008)
  [arXiv:0710.2067 [hep-ph]];
  Comput.\ Phys.\ Commun.\  {\bf 180}, 1579 (2009)
  [arXiv:0808.3144 [hep-ph]];
  Comput.\ Phys.\ Commun.\  {\bf 180}, 1718 (2009).

\bibitem{2HDMC}
  D.~Eriksson, J.~Rathsman and O.~Stal,
  Comput.\ Phys.\ Commun.\  {\bf 181}, 189 (2010)
  [arXiv:0902.0851 [hep-ph]];
  Comput.\ Phys.\ Commun.\  {\bf 181}, 833 (2010).
  
\bibitem{Misiak:2006zs} 
  M.~Misiak, H.~M.~Asatrian, K.~Bieri, M.~Czakon, A.~Czarnecki, T.~Ewerth, A.~Ferroglia and P.~Gambino {\it et al.},
  Phys.\ Rev.\ Lett.\  {\bf 98}, 022002 (2007)
  [hep-ph/0609232].

\bibitem{Li:2014fea} 
  X.-Q.~Li, J.~Lu and A.~Pich,
  JHEP {\bf 1406}, 022 (2014)
  [arXiv:1404.5865 [hep-ph]].
  
\bibitem{bsmmexp}
CMS and LHCb Collaborations,
 CMS-PAS-BPH-13-007, 
 available from \verb+http://cds.cern.ch+.
 
\bibitem{Djouadi:1995gt} 
  A.~Djouadi, M.~Spira and P.~M.~Zerwas,
  Z.\ Phys.\ C {\bf 70}, 427 (1996)
  [hep-ph/9511344].
  
\bibitem{Djouadi:1997yw} 
  A.~Djouadi, J.~Kalinowski and M.~Spira,
  Comput.\ Phys.\ Commun.\  {\bf 108}, 56 (1998)
  [hep-ph/9704448].
  
\bibitem{Djouadi:1995gv} 
  A.~Djouadi, J.~Kalinowski and P.~M.~Zerwas,
  Z.\ Phys.\ C {\bf 70}, 435 (1996)
  [hep-ph/9511342].
  
\bibitem{Akeroyd:1998dt} 
  A.~G.~Akeroyd,
  Nucl.\ Phys.\ B {\bf 544}, 557 (1999)
  [hep-ph/9806337].

\bibitem{HHG}
J.~F.~Gunion, H.~E.~Haber, G.~L.~Kane, and S.~Dawson, 
{\it The Higgs Hunter's Guide} (Westview, Boulder, Colorado, 2000).

  
\end{thebibliography}
\end{document}